\documentstyle[prl,aps,twocolumn,multicol,epsf]{revtex}
\tighten
\begin{document}   
\twocolumn[\hsize\textwidth\columnwidth\hsize\csname
@twocolumnfalse\endcsname
\title {Structure of the black hole's Cauchy horizon singularity}
\author{Lior M. Burko}
\address{
Department of Physics, Technion---Israel Institute of Technology,
32000 Haifa, Israel.}
\date{\today}

\maketitle

\begin{abstract}

We study the Cauchy horizon (CH) singularity of a
spherical charged black hole perturbed nonlinearly by a
self-gravitating massless scalar field. We show numerically
that the singularity is weak both at the early and at the late
sections of the CH,
where the focusing of the area coordinate $r$ is strong.
In the early section the metric is almost Reissner-Nordstr\"{o}m,
and the fields behave according to perturbation analysis.
We find exact analytical expressions for the gradients
of $r$ and of the scalar field, which are valid at both
sections. We then verify these analytical results numerically.
\newline
\newline
PACS number(s): 04.70.Bw, 04.20.Dw, 04.25.Dm
\end{abstract}


\vspace{3ex}
]






One of the long-standing interesting predictions of General Relativity (GR) 
is the occurrence of spacetime singularities inside black holes (BHs).
This issue is intriguing, because the laws of physics we currently understand 
(e.g. classical GR) are not valid at singularities, but some other, as yet unknown, 
laws take over from classical GR and control the structure of singularities.  
Thus, despite many efforts in the last few decades, 
the nature of the spacetime singularities
in a generic gravitational collapse---and, 
more generally, the final outcome of the collapse---are still open questions.

Until recently, the only known generic singularity was the BKL
singularity \cite{bkl}---an oscillatory spacelike singularity.
In the last few years, however, evidence has been steadily accumulating that 
another type of singularity forms at the Cauchy horizon (CH)
of spinning or charged BHs. The features of this new singularity
differ drastically from those of the previously-known
singularities like, e.g., Schwarzschild or BKL: First, the CH
singularity is null rather than spacelike \cite{pi,ori1,ori2}; Second,
it is weak \cite{ori1,ori2}. Namely, the tidal distortion experienced by
an infalling extended test body is finite (and, moreover, 
is typically negligibly small) as it hits
the singularity \cite{tipler}. Yet, curvature scalars diverge there \cite{pi,ori2} 
(in the spherical charged case, this is expressed by mass inflation \cite{pi}).

For uncharged spinning BHs (the more realistic case), 
the evidence in favor of this
new picture emerges primarily from a systematic linear and non-linear 
perturbative analysis
\cite{ori2,ori-grg}. In addition, the local existence and
genericity of a null weak singularity in solutions of the vacuum
Einstein equations was demonstrated in Ref. \cite{flanagan}. 
(The compliance of null weak singularities
with the constraint equations was demonstrated in \cite{bc}.)
In the case of a spherical charged BH, the weakness of the singularity 
was first demonstrated in \cite{ori1}.  
More recently, an approximate leading-order analysis \cite{israel}
confirmed the local consistency of this new picture. 

Despite these recent advances, our understanding of the null weak CH 
singularity is still far from being complete. 
In particular, it is important to verify this new picture by
performing independent, non-perturbative, analyses. This motivates 
one to employ numerical tools to
study the structure of the CH singularity. The numerical simulation of 
spinning BHs is difficult, as they are non-spherical. One is thus led
to study, numerically, the inner structure of a spherical charged 
BH; hopefully, it may serve as a useful toy model for a spinning BH.

The first numerical analysis of a perturbed charged BH's interior 
was carried out by Gnedin and Gnedin \cite{gnedin}, 
who analyzed the spherically-symmetric 
gravitational collapse of a self-gravitating scalar field over a charged 
background. 
The coordinates (and numerical grid) used there, however, do not allow 
getting even close to the CH.
More recently, Brady and Smith (BS) \cite{brady} numerically explored
the mass-inflation singularity inside a spherical charged BH perturbed 
nonlinearly by a scalar field. This analysis
confirmed several aspects of the above new picture: 
It demonstrated the existence of a null singularity
at the CH, where the mass function $m$ diverges but the radial Schwarzschild 
coordenate $r$ is nonzero. The quantity 
$r$ was found to decrease monotonically with increasing 
retarded time along the CH, due to the nonlinear focusing,
until it shrinks to zero (at which point the singularity becomes spacelike). 
It also provided evidence for the weakness of the singularity. 
Despite its remarkable achievements, however,
this analysis left one important issue unresolved:  
To what extent is the  
perturbative approach applicable at (and near) the CH singularity? 
BS reported on an inconsistency with the predictions of perturbation analysis,
manifested by the non-zero value of $\sigma$ (see \cite{brady}), namely a 
finite deviation of the power-law indices from the integer values predicted 
by perturbative analyses. 
This issue is crucial, because for realistic (i.e., spinning and uncharged) 
black holes 
the only direct evidence at present for the actual occurrence of a null weak 
singularity stems from the perturbative analysis \cite{ori2}. A failure of 
the perturbative approach in the spherical charged case would therefore cast 
doubts on our understanding of realistic black holes' interiors. 

In this paper we report on a numerical and analytical investigation, 
which we carried out in order to answer this and other questions. We consider
the model of a spherical charged black hole non-linearly perturbed by a 
spherically-symmetric, self-gravitating, neutral,
massless scalar field $\Phi$ (the same model as in \cite{gnedin,brady}). 
We shall first present our numerical results which show 
that the CH singularity
is indeed weak---not only in the early part of the CH, but also all the 
way down to the point of
complete focusing ($r=0$). Then, we study the asymptotic 
behavior of perturbations at the early part
of the CH, and demonstrate the full compliance with the 
predictions of perturbation theory
(in particular, we find that $\sigma\equiv 0$).
In addition, we shall show that despite the divergence of $m$, the metric
functions at the asymptotically-early part of the CH are remarkably close to the
unperturbed Reissner-Nordstr\"{o}m (RN) metric functions---which is again a
prediction of the perturbation analysis (according to the latter, 
the metric perturbations should
vanish at the asymptotic past ``edge'' of the CH, despite the divergence of 
the curvature \cite{ori2}).
Then, we shall analyze the behavior of the blue-shift factors $r_{,v}$ 
and $\Phi_{,v}$, as a function of $u$, at
the late (i.e. strong-focusing) part of the CH, 
where $u$ and $v$ are ingoing and
outgoing null coordinates, respectively (see below).
The asymptotic behavior of $r_{,v}$ and $\Phi_{,v}$ is essential, because
it is primarily these entities that are responsible to the divergence of 
curvature at the CH.
We shall present exact analytic expressions for these entities,
and verify them numerically. The expression we obtain for $\Phi_{,v}$, in 
addition to our numerical results, 
shows that $\sigma$ vanishes not only at the early part of the CH, but also 
everywhere along it.

We write the general spherically-symmetric line element in double-null coordinates,
\begin{equation}
\,ds ^{2}=-f(u,v)\,du\,dv+r^{2}(u,v)\,d\Omega^{2},
\label{metric}
\end{equation}
where $\,d\Omega^{2}$ is the unit two-sphere. As the source term for the
Einstein equations we take the contributions of both the
scalar field and the (sourceless) spherically-symmetric electric field
(see \cite{burko-ori} for more details). The dynamical field equations are  
\begin{eqnarray}
\Phi_{,uv}+\left(r_{,u}\Phi_{,v}+r_{,v}\Phi_{,u}\right)/r=0
\label{KGEQ}
\end{eqnarray}
\begin{eqnarray}
f_{,uv}&=&\frac{f_{,u}f_{,v}}{f}+f\left\{ \frac{1}{2r^{2}}\left[4r_{,u}r_{,v}+
f\left( 1-2\frac{Q^{2}}{r^{2}}\right)\right]\right.\nonumber \\
&-&\left. 2\Phi_{,u}\Phi_{,v} \right\}
\label{EEQ2}
\end{eqnarray}
\begin{eqnarray}
r_{,uv}=-\frac{r_{,u}r_{,v}}{r}-\frac{f}{4r}\left(1-\frac{Q^{2}}
{r^{2}}\right),
\label{EEQ1}
\end{eqnarray}
where the constant $Q$ is the electric charge.
Equations  (\ref{KGEQ})--(\ref{EEQ1}) are 
supplemented by two constraint equations:
\begin{eqnarray}
r_{,uu}-(\ln f)_{,u}r_{,u}+r(\Phi_{,u})^{2}=0
\label{con1}
\end{eqnarray}
\begin{eqnarray}
r_{,vv}-(\ln f)_{,v}r_{,v}+r(\Phi_{,v})^{2}=0.
\label{con2}
\end{eqnarray}
\begin{figure}
\epsfxsize=8.0cm
\epsffile{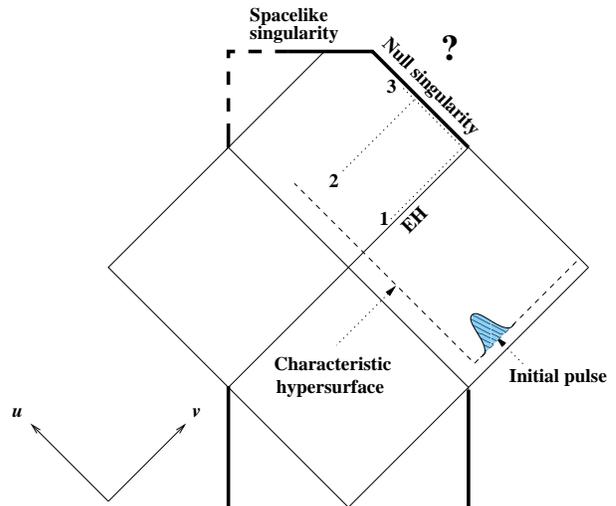}
\caption{The Penrose diagram of the simulated spacetime. 
Singularities are displayed by thick lines, and the characteristic hypersurface 
on which the initial data are specified are marked with dashed lines.  
The various fields are probed along outgoing (such as 1---in the early 
sections---and 2---in the late sections) and 
ingoing (such as 3) null rays. It is still unknown whether there is a continuation 
of the spacetime manifold beyond the CH.  
Our domain of integration cannot include the entire 
spacelike portion of the singularity. These uncovered areas 
are marked by dashed thick lines. }
\label{fig1}
\end{figure}
It turns out that it is advantageous to substitute $f(u,v)=2e^{2 s(u,v)}$ 
for the numerical integration near the CH.  
Our initial-value setup is described in Ref. \cite{burko-ori}: 
The geometry is initially RN,
with initial mass $M_0=1$ and charge $Q$, and no scalar field.
At some moment $v$, however, it is modified by an ingoing 
scalar-field pulse of
a squared-sine shape with amplitude $A$. $\Phi$ vanishes everywhere on 
the initial surface except in a finite range $v_1<v<v_2$. 
The results presented below relate to $v_1=10$, $v_2=20$, 
$Q=0.95$ and $A=8\times 10^{-2}$, 
unless stated otherwise.  
In this case, due to the scalar-field energy, the BH's external mass
approaches the final mass $M_f \approx 1.4$. 
(We also checked other values of $0.5<Q/M_{f}<.99$ and $A$, 
and obtained similar results.)
Note that our outgoing initial null hypersurface is located 
outside the event horizon (EH) (unlike in Ref. \cite{brady}). 
Therefore, we do not have to make any assumption about
the inverse power tails at the EH; these are created automatically by the 
dynamical evolution.  
Our numerical scheme is essentially the same as described in Ref. \cite{burko-ori}
(there are few modifications, which will be described elsewhere \cite{long}): 
It is based on
free evolution in double-null coordinates.
The code is stable and second-order accurate \cite{burko-ori}. Our numerical 
setup is displayed in Figure 1, imbedded in the Penrose diagram of the 
simulated spacetime. 

The null coordinates $u$ and $v$ are defined in Ref. \cite{burko-ori}: 
They are taken to be
linear with $r$ on the two characteristic initial segments.
For the presentation and interpretation of the results,
we shall occasionally use other types of double-null coordinates:
Eddington-like coordinates $u_e$ and $v_e$, and Kruskal-like
coordinates (of the {\it inner} horizon of RN) $U$ and $V$.
(In the perturbed spacetime, the Eddington-like
and Kruskal-like coordinates are defined with respect to the 
asymptotic ``vertex'' region,
where the metric perturbations vanish asymptotically; see \cite{long})
Recall that at $v\gg M_{f}$, $v$ is closely related to $v_e$ \cite{burko-ori}.

Our numerical simulations confirm the presence of a null 
singularity at the CH, where
$m$ diverges and $r$ is nonzero. Along the CH singularity, $r$
decreases monotonically, until it shrinks to zero, at which point the 
singularity becomes spacelike.
This situation was already found numerically by BS \cite{brady}.

\begin{figure}
\epsfxsize=8.0cm
\epsffile{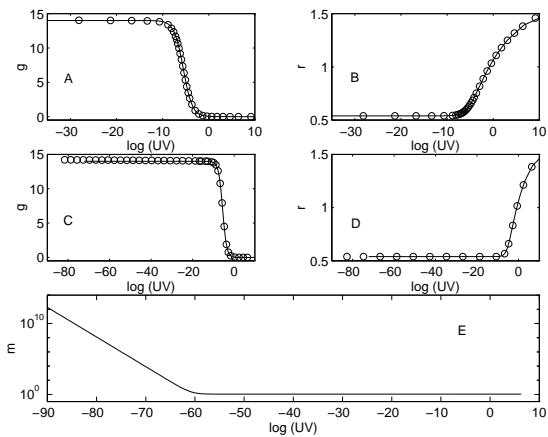}
\caption{Metric functions in the early sections of the CH. 
$g$ [(A) and (B)] and $r$ 
[(C) and (D)] as functions of $\log (UV)$ along an outgoing ray and
an ingoing ray at $v=80$, respectively.
The solid lines are the RN values (computed according to the BH's external
parameters at $v=75$), and the circles represent the numerical 
values. (E): The mass function along the same outgoing ray as (A) and (C). 
Here we took $N=80$ and $A=5\times 10^{-2}$, which corresponds 
to $M_{f}\approx 1.1$.}
\label{fig2}
\end{figure}

We shall first discuss the early part of the CH singularity, i.e., 
the part where the focusing of
$r$ is still negligible. Our first goal is to demonstrate that the singularity is weak.
In terms of the double-null metric (\ref{metric}), the singularity will be weak if
coordinates $\hat u(u),\hat v(v)$ can be chosen such that
both $r$ and $g_{\hat u \hat v}$ are finite and
nonzero at the CH. 
The numerical analysis by BS already demonstrated the finiteness of $r$, which 
we recover in our results. (Note that $r$ is independent of the choice of the 
null coordinates.) Figure 2(A) 
displays the metric function  $g\equiv -2g_{\hat u \hat v}$ in Kruskal-like 
coordinates $U,V$ along an outgoing null ray 
that intersects the early section of the CH singularity. 
The CH is located at $V=0$ (corresponding
to $v\to\infty$). This figure clearly demonstrates the finiteness of $g_{UV}$,
from which the weakness of the singularity follows.

The perturbation analysis also predicts that both
the scalar field and the metric perturbations will be 
arbitrarily small at the early section of the
CH. In other words, both metric functions $r$ and $g$ 
should be arbitrarily close to
the corresponding RN metric functions. 
This behavior is indeed demonstrated in Fig. 1.
In this figure, $g$ and $r$ are displayed along lines $u={\rm const}$ 
[Figs. 2(A) and 2(B)]
and $v={\rm const}$ [Figs. 2(C) and 2(D)]. 
The similarity of the analytic RN functions and the
numerically-obtained functions of the perturbed spacetime is remarkable. 
(We emphasize, though, that despite the similarity in the values of the metric 
functions to RN, our geometry is drastically 
different from that of RN, in the sense 
that in our case curvature blows up at the CH.) 

\begin{figure}
\epsfxsize=8.0cm
\epsffile{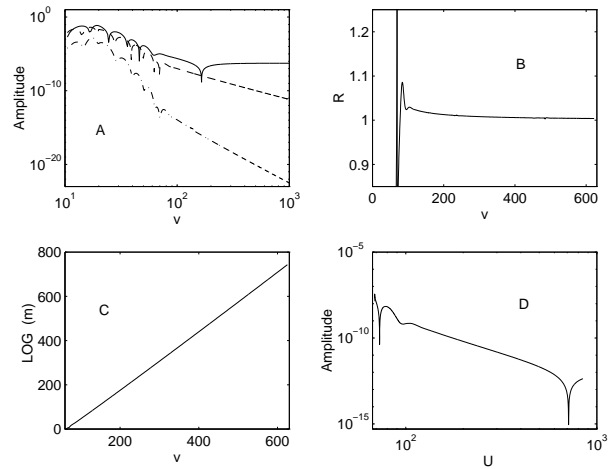}
\caption{Fields in the early sections of the CH. (A):
$\Psi$ (solid), $\Psi_{,v}$ (dashed), and $r_{,v}$ (dash-dotted) along
an outgoing ray, as a function of $v$. (B): $R$, and (C): $\log m$, 
along the same outgoing ray, as a function of  $v$. 
(D): $\Psi$ along an ingoing ray, as
a function of $U$.}
\label{fig3}
\end{figure}

The validity of the perturbation analysis is checked in more detail in 
Fig. 3, which displays
the power-law behavior of $\Psi\equiv r\Phi$, $\Psi_{,v}$, and $r_{,v}$ near the early
section of the CH. 
Fig. 3(A) displays these entities along an outgoing ray. Both the ringing 
and the power-law tails
are apparent (the graph of $\Psi$ approaches a nonzero limiting value, due to the finite value
of $U$ on that ray).
Fig. 3(B) displays $R\equiv \Psi_{,v}/\Psi_{,v}^{\rm lin}$, where
$\Psi_{,v}^{\rm lin}=\Psi_{,v}^{\rm EH}[2(M_f/Q)^{2}-1]$ 
is the asymptotic form of
$\Psi_{,v}$ as predicted by linear perturbation theory 
\cite{gursel}, and $\Psi_{,v}^{\rm EH}$
is the ($v$-dependent) value of $\Psi_{,v}$ at the EH.
Numerically, we find that at large $v$, $R$ tends asymptotically to 
$1$---it deviates from unity by no more
than $2\times 10^{-3}$ at $v=600$. A similar result was obtained for various values
of $Q$ and $A$. We thus find no evidence for charge-dependent internal power-law
indices, and we conclude that the parameter $\sigma$ of Ref. \cite{brady} vanishes
identically. The behavior
of $m$ along the same outgoing ray is shown in Fig. 3(C). Clearly,
the exponential growth of $m$ does not affect the validity of the
perturbation analysis. Finally, Fig. 3(D) displays $\Psi$ along an {\it 
ingoing} null ray at 
very large $v$. The local power index (see \cite{burko-ori}) of $u_e$ is found numerically
to be $3.1$, in a remarkable agreement with the value $3$ predicted by linear perturbation theory.
We conclude that the linear perturbation analysis describes the evolution 
of perturbation
near the vertex well, despite the divergence of $m$ (and the curvature).
This confirms the prediction of nonlinear perturbation analysis \cite{ori2}, namely
that nonlinear perturbations will be negligible at the
early part of the CH, compared to the linear perturbations.

We turn now to explore the late part of the CH singularity, where 
focusing is strong.
First, we numerically verify the weakness of the singularity in this part too.
Figure 4(A) shows $G\equiv -2g_{uV}$ along an outgoing null ray in the late
part of the CH, where the value of $r$ has shrunk to  $10\%$ of the value it 
had when the CH was first formed.  
We present here the results for various values of the grid-parameter
$N$ (see \cite{burko-ori}), in order to demonstrate the second-order 
numerical convergence.
$g_{uV}$ approaches a finite value at the CH ($v\to \infty$). 
At the same time, 
the mass function (and curvature) grows exponentially with $v$ 
[see Fig. 4(B)]. We conclude
that the entire null CH singularity is weak, 
even at the region of strong focusing.

\begin{figure}
\epsfxsize=8.0cm
\epsffile{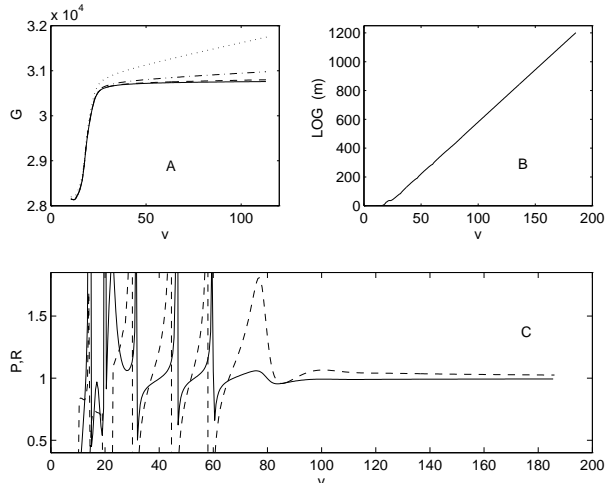}
\caption{The nonlinear regime
(contraction of the CH of $90\%$): (A) $G$
along an outgoing null ray, as a function of $v$.
Dotted: $N=20$, dash-dotted: $N=40$, dashed: $N=80$,
and solid line: $N=160$. (B) $\log m$ along the
same outgoing ray. (C): 
$R$ (dotted) and $P$ (solid) 
along the same outgoing null ray,
as a function of $v$.}
\label{fig2a}
\end{figure}

Next we study, analytically, the behavior of the blue-shift factors 
$r_{,v}$ and $\Phi_{,v}$
along the contracting CH. 
Here, we shall present the results; the full derivation will be
presented in Ref. \cite{long}. The field equations (\ref{KGEQ}) and (\ref{EEQ1}) can 
be integrated exactly along the CH singularity \cite{long}. For 
$r_{,v}$ we find
\begin{eqnarray}
{\left(r^{2}\right)}_{,v_{e}}=-(1/\kappa_{-})\;{\Psi_{,v_{e}}}^{2},
\label{lin3}
\end{eqnarray}
and for $\Psi_{,v}$ we find
\begin{eqnarray}
\Psi_{,v}=\left[2\left(M_f/Q\right)-1\right]\;\Psi_{,v}^{\rm{EH}}.
\label{lin30}
\end{eqnarray}
Here, $\kappa_{-}$ is the surface gravity at the RN inner horizon with parameters 
$M_f$ and $Q$. 
For convenience, Eqs. (\ref{lin3}) and (\ref{lin30}) are expressed in terms of $\Psi$. 
Note that (\ref{lin30}) is invariant to a gauge transformation $v\to \tilde{v}(v)$, 
whereas (\ref{lin3}) is not. [Eq. (\ref{lin3}) refers explicitly to the 
derivatives with respect to the Eddington-like coordinate $v_e$.] 

In order to verify this prediction, we calculated numerically 
$R\equiv \Psi_{,v}/\{\Psi_{,v}^{\rm EH}[2(M_f/Q)^{2}-1]\}$ 
and $P\equiv -{\left(r^{2}\right)}_{,v_{e}}/\left[ 
(1/\kappa_{-}){\Psi_{,v_{e}}}^{2}\right]$, as functions of $v$,
along an outgoing ray located at a region of $90\%$ focusing of the CH.
The results, presented in Fig. 4(C), are in
excellent agreement with the above theoretical prediction, $R=1=P$.

To summarize, we have confirmed, numerically (for a spherical charged 
black hole) the main
predictions of the perturbation analysis 
(which apply both to spherical charged and nonspherical, spinning BHs): The null
singularity at the CH is found to be weak.
In the asymptotic early section of the CH the metric functions 
approach arbitrarily close
to the corresponding metric functions in RN, and, moreover, 
the perturbations are well described
by the {\it linear} perturbation analysis. This confirms the conclusions of the perturbation
analysis \cite{ori2}, that the non-linear effects are negligible at the early section of 
the CH singularity.
In addition, we analytically derived exact expressions for the diverging blue-shift
factors $r_{,v}$ and $\Phi_{,v}$, which are valid everywhere along the contracting CH.

I am indebted to Amos Ori for numerous invaluable discussions and useful 
comments.


\end{document}